\documentclass[aps,prb,twocolumn,superscriptaddress,showpacs,showkeys,floatfix]{revtex4-1}

\usepackage{amsmath,latexsym,amsfonts,amssymb}
\usepackage{natbib}
\usepackage{epsfig}
\usepackage{graphicx}
\usepackage{inputenc}
\usepackage{color}
\usepackage{setspace}
\usepackage{multirow}
\usepackage{array}
\usepackage{tabularx}
\usepackage{bm}
\usepackage{color}
\usepackage{gensymb}
\usepackage{xcolor}
\usepackage[left]{lineno}
\usepackage{float} 
\usepackage{verbatim}

\newcommand{\poubelle}[1]{}
\newcommand{\gd}{GdMn$_2$O$_5$  }

\begin{document}

%=============================================================================

%\title{Magnetic field induced novel multiferroic phase in Gd$Mn_2$O$_5$}
\title{Electronic ground-state hysteresis under magnetic field in \gd\ }

\author{V. Balédent}
\affiliation{Universit\'e Paris-Saclay, CNRS, Laboratoire de Physique des Solides, 91405, Orsay, France.}
\email [Corresponding author: ] {victor.baledent@universite-paris-saclay.fr}

\author{A. Vaunat}
\affiliation{Universit\'e Paris-Saclay, CNRS, Laboratoire de Physique des Solides, 91405, Orsay, France.}
\affiliation{Synchrotron SOLEIL, L'Orme des Merisiers, Saint Aubin BP 48, 91192, Gif-sur-Yvette, France}
\affiliation{Laboratoire L\'eon Brillouin, CEA, CNRS, Universit\'e Paris-Saclay, 91191, Gif sur Yvette, France}

\author{S. Petit}
\affiliation{Laboratoire L\'eon Brillouin, CEA, CNRS, Universit\'e Paris-Saclay, 91191, Gif sur Yvette, France}

\author{L. Nataf}
\affiliation{Synchrotron SOLEIL, L'Orme des Merisiers, Saint Aubin BP 48, 91192, Gif-sur-Yvette, France}

\author{S. Chattopadhyay}
\affiliation{Dresden High Magnetic Field Laboratory (HLD-EMFL), Helmholtz-Zentrum Dresden-Rossendorf, 01328 Dresden, Germany}
\affiliation{UGC-DAE Consortium for Scientific Research Mumbai Centre, 246-C CFB, BARC Campus, Mumbai 400085, India}

\author {S. Raymond}
\affiliation{Universit\'e. Grenoble Alpes, CEA, IRIG, MEM, MDN, 38000 Grenoble, France} 

\author{P. Foury-Leylekian}
\affiliation{Universit\'e Paris-Saclay, CNRS, Laboratoire de Physique des Solides, 91405, Orsay, France.}

\date{\today}

\begin{abstract}

In this paper, we investigate the physical properties of the type II multiferroic \gd\  material by means of neutrons scattering, electric polarization and magnetization measurements. A complex (T,H) phase diagram shows up, with especially a field induced magnetic transition around 12 T at low temperature. The high field phase is accompanied by an additional electric polarization along both the $a$ and $b$ directions, as authorized by symmetry, but never observed experimentally up to now. While the magnetic properties recover their initial states after driving the field back to zero, the polarization along $a$ shows a significant increase. This behavior is observed for all directions of the magnetic field. It constitutes a novel and striking manifestation of the magneto-electric coupling, resulting in the establishment of a new ground state at zero magnetic field. 
\end{abstract}
%============================================================================================================================

\maketitle{}
\section{Introduction}

The search for new materials with remarkable properties is a major concern for many condensed matter physicists. In this quest, a simple idea consists in combining, within the same material, two different properties, even those which are {\it  a priori} mutually exclusive. Magneto-electrical multiferroics are one of the outstanding examples. These materials allow for the manipulation of the magnetic state using an electric field via the coupling between ferroelectricity and magnetism, which represents a great potential in the field of spintronics or information storage. Different routes have been explored to obtain such properties. Artificial materials, such as heterostructures \cite{Garcia2015} for instance, consist in alternating layers with different properties. The coupling is then induced by proximity effect. Another extensively studied route focuses on materials whose magnetic and ferroelectric properties coexist in the bulk. Even if quite difficult to find, their list is getting longer and longer. If ferroelectricity and magnetism have a distinct microscopic origin, these materials are called type I multiferroics. One of the main interests of this family is to offer multiferroic properties at room temperature, as in BiFeO$_3$, but with the disadvantage of a weak coupling inherent to the distinct origin of the two orders. In type II multiferroics, ferroelectricity is induced by magnetism, the two orders are then intrinsically strongly coupled, as in RMn$_2$O$_5$ manganites for instance. This last variety of compounds has received much attention because of the fundamental problem posed by the origin of this intrinsic coupling, which results in a complex ground state where lattice, electronic and magnetic degrees of freedom are entangled.\\

In this study, we report the possibility of manipulating the electronic ground state of a multiferroic material in a non-reversible manner using a magnetic field. We show that a first order magnetic transition occurs at 12 T, involving both charge and spin degrees of freedom. While the electronic properties exhibit significant hysteresis, the magnetism exhibits only a weak 1 T wide loop. This opens up a new avenue for creating a new functional ground state in multiferroic materials.

The type II RMn$_2$O$_5$ multiferroic family attracted attention for multiple reasons. First, while the average space group is $Pbam$ with a=7.2931 \AA, b= 8.5025 \AA, c=5.6743 \AA, this family crystallizes in a non-centrosymmetric structure at room temperature \cite{Baledent2015}. The actual space group $Pm$, with the mirror perpendicular to the $c$ axis of $Pbam$ effectively allows for electrical polarization in the $(a,b)$ plane. While such room temperature polarization has not been confirmed experimentally, it has been calculated using DFT \cite{Dai2020}. This polarization shows two components, electronic and ionic, whose amplitudes depend on the path followed in the temperature-electric field phase diagram. Second, one of the magnetic transitions, below T$_N$= 33K, is accompanied by the rise of an additional contribution along the $b$ direction \cite{Lee2013}, that adds to the initial polarization. It has been proven experimentally that the exchange-striction mechanism is at play in this family \cite{Yahia2018}. This additional spin-induced contribution to the polarization is maximized in \gd, where it culminates at 360 nC.cm$^{-2}$. Third, magnetic field has proven very efficient in modifying the electronic ground state in the whole RMn$_2$O$_5$ family. Indeed, a modulation of polarization has been observed below 2 T in \gd\  \cite{Lee2013} and the emergence of a polarization together with ferromagnetism has been reported in PrMn$_2$O$_5$ above 15 T \cite{Chattopadhyay2020}. All these reasons motivate the present high magnetic field study of \gd\ . 

The possibility to modify electronic properties with magnetic field along $a$ has been reported recently \cite{Ponet2022}. The authors propose that the magnetic field can induce a transition between four different topological magnetoelectric states, formed by the two zero field states and the two high field ones. The orientation of the magnetic field is also important. For a field purely along the $a$ direction, the final state is identical to the initial state, the polarization along $b$ is unchanged and the magnetic state either. For a field at 10 degrees from the $a$-axis in the $(a,b)$ plane, the initial and final states however are different, changing the polarization along the $b$-axis. The magnetic structure is also modified, but is experimentally indistinguishable from the initial structure: there are two identical degenerate states respecting the very same symmetries and magnetic space group. The existence of these two equivalent magnetic structures at zero field giving rise to different polarizations had already been pointed out by DFT calculations\cite{Dai2020}. The possibility to switch from one to the other using a magnetic field along the $a$-axis to modify the polarization along the $b$-axis seems to be linked to the topological properties of these states. We present here a combination of magnetization, electric polarization and neutron diffraction to investigate further this mechanism.

\section{Results}

%Simulation of magnetic structure ?
% Confirmed by magnetisation + field direction
%===========================================================================================================================
\begin{figure}[ht]
	\includegraphics [width=0.8\linewidth, angle=0]{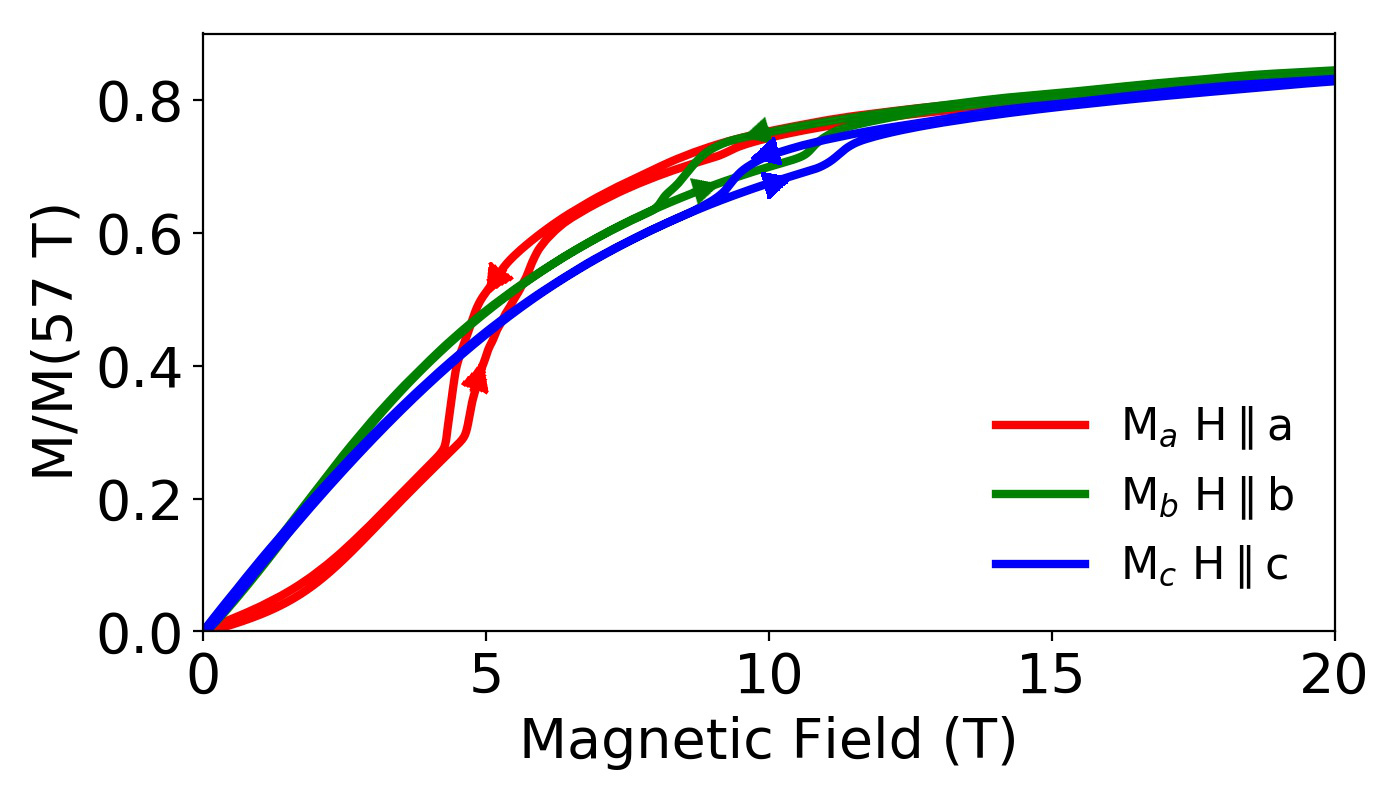}
	\caption{Magnetization along the three different crystallographic axes as a function of magnetic field at T=2 K.}
	\label{M_H}
\end{figure}  
%===========================================================================================================================

\subsection{Magnetization}
We performed magnetization measurements with the magnetic field applied along the three crystallographic axes of $GdMn_2O_5$. These field-sweep measurements were carried out up to $\sim$57 T at 2 K using pulsed magnetic fields available at the Dresden High Magnetic Field Laboratory (HLD-EMFL). The duration of each pulse was $\sim$20 ms. Reproducibility of the data was ensured by repeating the measurements for each direction. As can be observed in Fig. \ref{M_H}, the magnetization along the $b$ direction shows a clear hysteresis loop around 10 T. Such hysteresis is also visible for the field along $a$ and $c$ around 5 and 11 T respectively. This corroborates the results obtained for the $a$ direction in ref \cite{Lee2013}.

%Polarization
\subsection{Electric polarization}

%===========================================================================================================================
\begin{figure}[ht]
	\includegraphics [width=0.8\linewidth, angle=0]{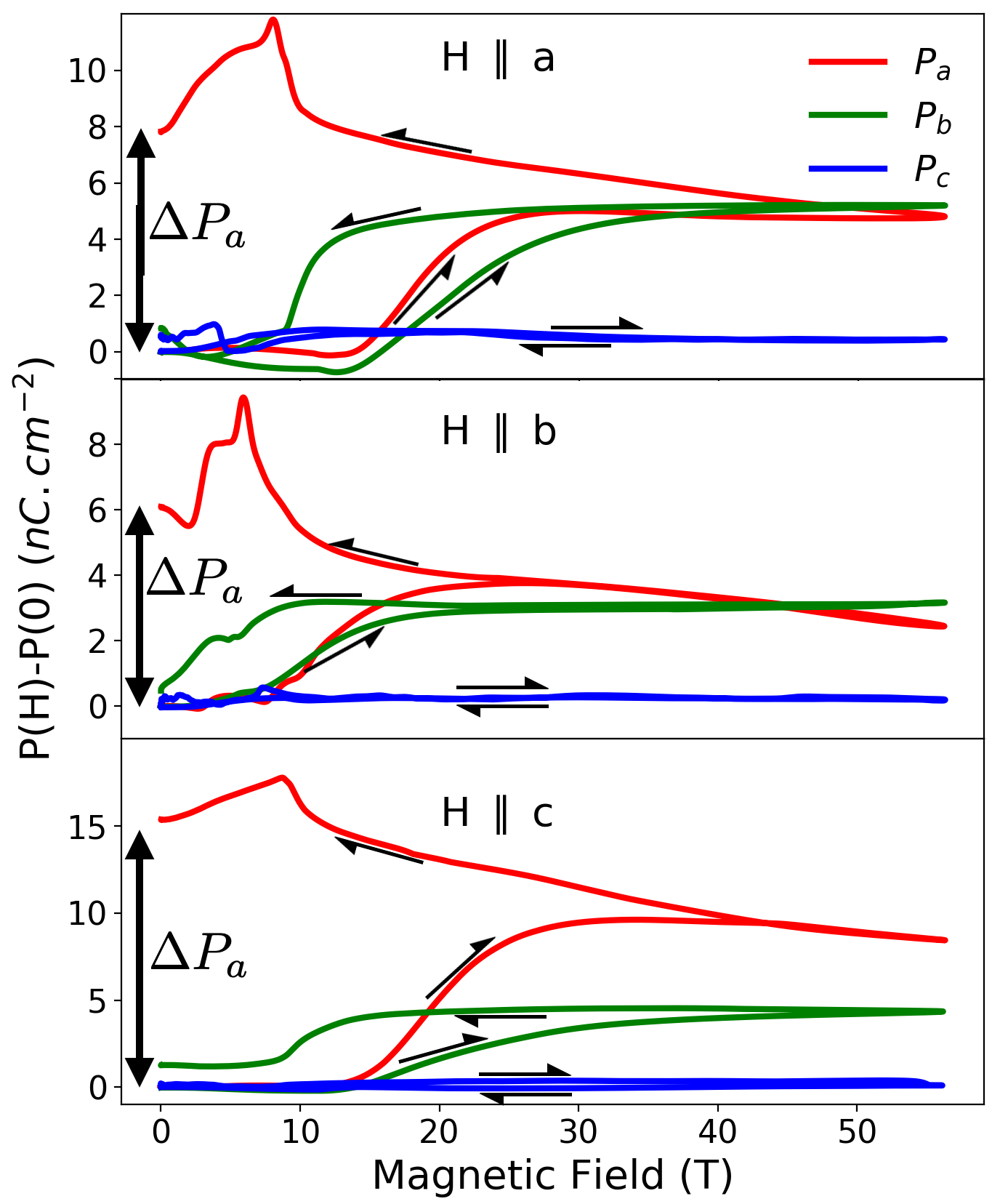}
	\caption{Variation of the electric polarization along $a$, $b$ and $c$ with respect to zero field initial state as a function of increasing and decreasing magnetic field at 2 K. $\Delta P_a$ at zero field after field cycling is indicated for each direction of the magnetic field. Its average among the 3 orientations is around 10 nC.cm$^{-1}$  }
	\label{P_H}
\end{figure}  
%===========================================================================================================================

Motivated by the strong magneto-electric coupling in this material, we turned to the evolution of the electric polarization $P$ under magnetic field. Similar to the magnetization, the measurements were also performed up to $\sim$57 T using a pyroelectric technique \cite{Chattopadhyay2020} at the HLD with a pulse duration of $\sim$20 ms. Both sides of the $a$, $b$, and $c$ thinned samples were covered using silver paste with gold contact wires attached to each of the sides. For each direction of the magnetic field we measured the components of the electric polarization along the 3 crystallographic directions : in total we performed 9 configurations at 2 K. The field variations of the pyroelectric current $I$ were recorded by measuring the voltage variation across a 1 M$\Omega$ shunt resistor connected in series with the measurement circuit by a digital oscilloscope (Yokogawa DL750). To calculate the field dependent electric polarization for each of the nine configurations, we integrated the $I$($H$) data. All the measurements were repeated to verify the reproducibility of the data. \\

Results are shown in Fig. \ref{P_H}. First, regardless of the direction and intensity of the magnetic field, no polarization has been measured along the $c$ direction ($P_c \equiv 0$, blue curves). The observed small variations are most likely due to imperfect alignment of the crystal and the electrical contacts. As expected owing to the $Pm$ space group with the mirror perpendicular to $c$ axis, the electric polarization lies within the $(a,b)$ plane ($P_a$ and $P_b \ne 0$). $P_b$ (green curve), however shows a step-like increase, above 12, 10 and 15 T for increasing field along $a$, $b$ and $c$ respectively. Decreasing the field from 57 T, the polarization goes back to its initial value with a hysteresis loop. \\

For H along $b$, this behavior is similar to the one observed in magnetization at the same critical field around 10 T. Hence, the high field magnetic phase induces an additional ferroelectric component $P_b=$4 nC.cm$^{-2}$, on top of the one initially present at zero field (360 nC.cm$^{-2}$). Interestingly, such increase in $P_b$ is also visible for magnetic fields along $a$ and $c$. \\
It's worth noting that only $P_b$ polarization has been experimentally measured at low temperature in the $CM1$ phase. Although the $Pm$ space group also allows non-zero $P_a$ polarization, this has never been demonstrated experimentally. Applying a magnetic field changes the game : as can be seen in Fig. \ref{P_H}, $P_a$ increases with increasing the field, similarly to $P_b$. Even more astonishing, with decreasing the field back to zero, $P_a$ does not go back to zero, but shows a residual polarization $\Delta P_a$ between 6 and 15 nC.cm$^{-2}$ at 0 T depending on the field orientation and thus around 10 nC.cm$^{-1}$ in average.\\

\subsection{Neutron scattering}
\begin{figure}[ht]
	\includegraphics [width=0.99\linewidth, angle=0]{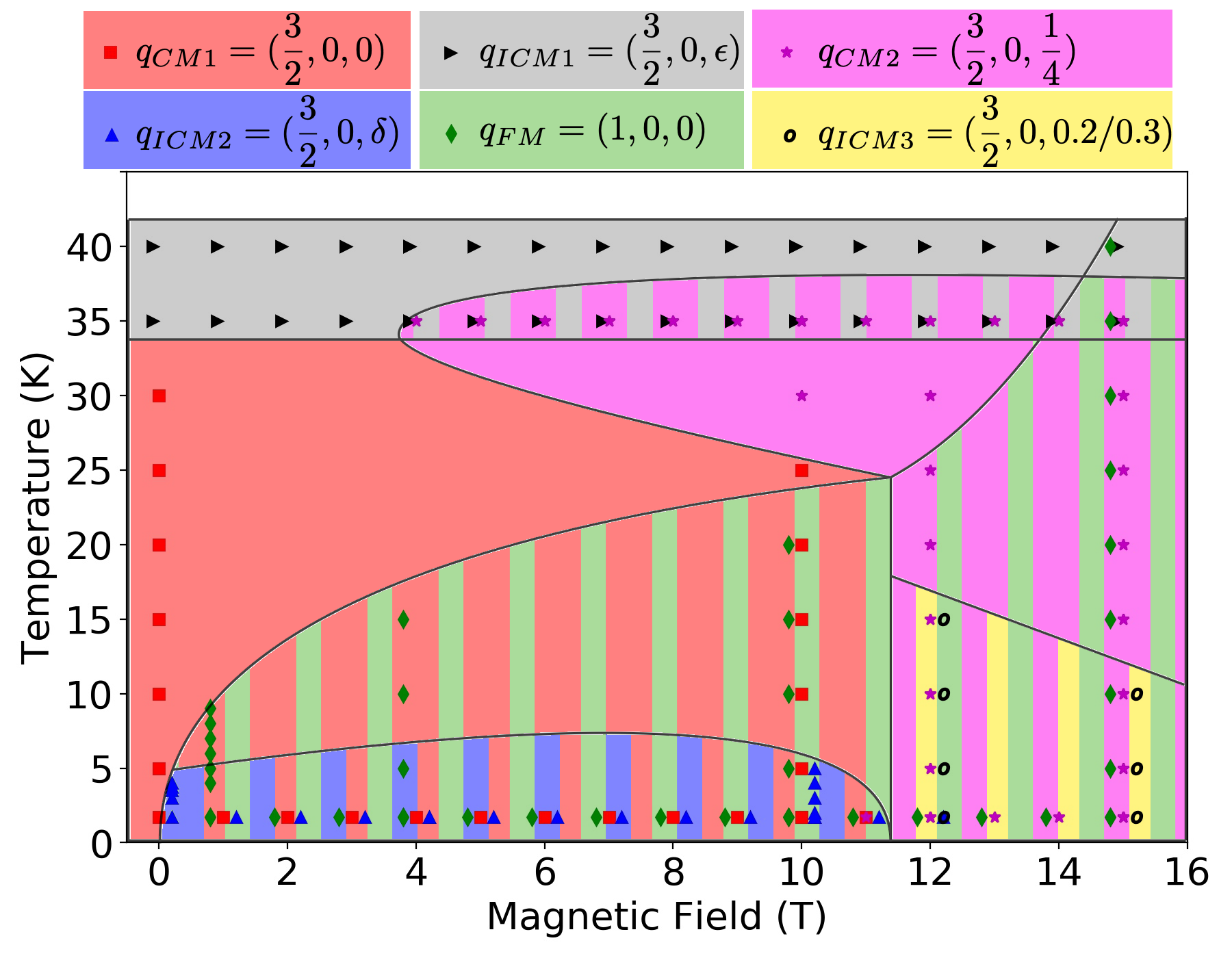}
	\caption{Schematic temperature-magnetic field phase diagram deduced from the neutron diffraction measurements. Points represents the region in the phase diagram where a magnetic signal has been observed at the different wavevectors indicated in the labels. The presence of multiple colored stripes denotes the coexistence of several magnetic peaks with different propagation vectors.}
	\label{PhaseDiagramm}
\end{figure}

In order to investigate further the magnetic properties under magnetic field of GdMn$_2$O$_5$, the temperature-magnetic field phase diagram was investigated by means of single crystal neutron diffraction. The experiment was carried out on the IN12 cold triple axis spectrometer, a CEA-Juelich CRG installed at ILL (Grenoble, France). Since the magnetic propagation vector is $(0.5, 0, 0)$ or $(0.5, 0, \delta)$ depending on the temperature in all members of the RMn$_2$O$_5$ family, the natural scattering plane is the $(a,c)$ plane, allowing access to magnetic peaks of the form $(H,0,L)$. The magnetic field was applied along the vertical direction, hence the $b$ direction. The evolution of selected reciprocal space regions was then recorded as a function of magnetic field and temperature, leading to the schematic phase diagram shown in Fig. \ref{PhaseDiagramm}. Several structures appear, characterized by different Bragg reflections and that we shall put in different categories:
\begin{itemize}
\item $q_{ICM1}=(3/2,0,\epsilon \approx 0.19)$  (gray), $q_{CM1}=(3/2,0,0)$ (red) and $q_{ICM2}=(3/2,0,\delta \approx 0.43)$ (blue) are characteristic of the zero field, high, intermediate and low temperature structures respectively, already reported in literature \cite{Yahia2018}. The incommensurability along the $c$ axis indicates that the magnetic moments wrap around the $c$ axis, forming a helicoidal type structure, and reflecting an exchange frustration along this direction. It is worth mentioning that the $q_{CM1}$ phase is fragile: strikingly, the spin wave dispersion, measured on the very same crystal in the 5 to 30 K temperature range \cite{Vaunat2021}, does not go soft at $q_{CM1}$ but at $q_{ICM2}$, indicating that the incommensurate phase is the incipient ground state.

\item $q_{FM}=(1,0,0)$ (green) reflects a ferromagnetic component, imposed by the field, with magnetic moments also along the applied magnetic field. 

\item $q_{CM2}=(3/2,0,1/4)$ (magenta) and $q_{ICM3}=(3/2,0,0.2-0.3)$ (yellow) are characteristic of field induced high and low temperature structures respectively. The magnetic structure inferred from $q_{CM2}$ is a quadrupling of the $q_{CM1}$ magnetic unit cell along the $c$-axis, and can also be seen as a commensurate lock-in of the periodicity along the same direction with respect to the $q_{ICM2}$ phase. Interestingly, this $q_{CM2}$ magnetic propagation vector corresponds to the ground state of the low temperature magnetic phase observed in numerous other members of the family (R=Tb, Dy, Ho, Er, Tm and Yb). \\
\end{itemize}

\begin{figure*}[!t]
	\includegraphics [width=0.99\linewidth, angle=0]{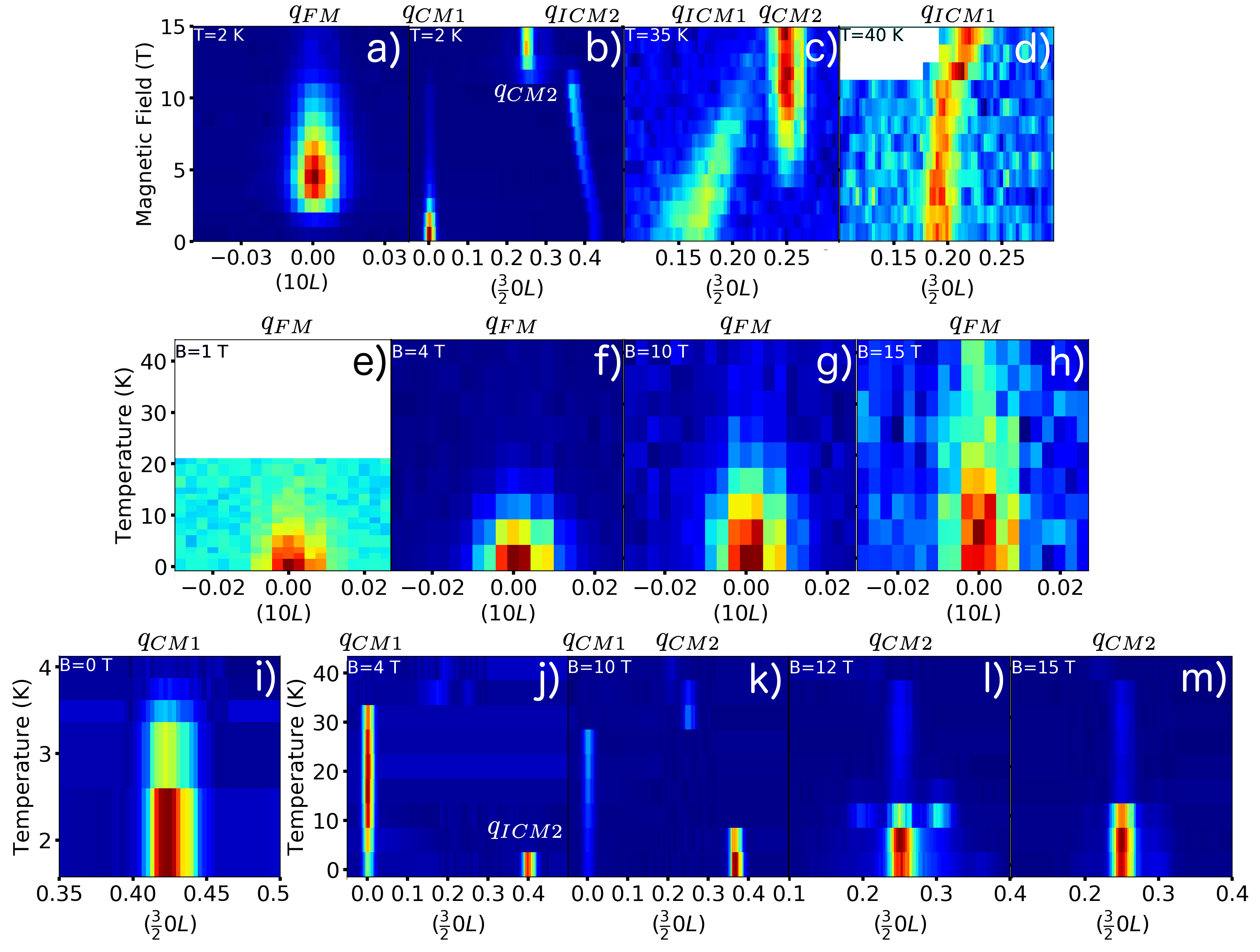}
	\caption{Color map of the magnetic intensity on the (1,0,L) and (1.5,0,L) Bragg reflections as a function of temperature and magnetic field. Color scales are not comparable to each other.}
	\label{Maps}
\end{figure*}

\begin{figure}[ht]
	\includegraphics [width=0.7\linewidth, angle=0]{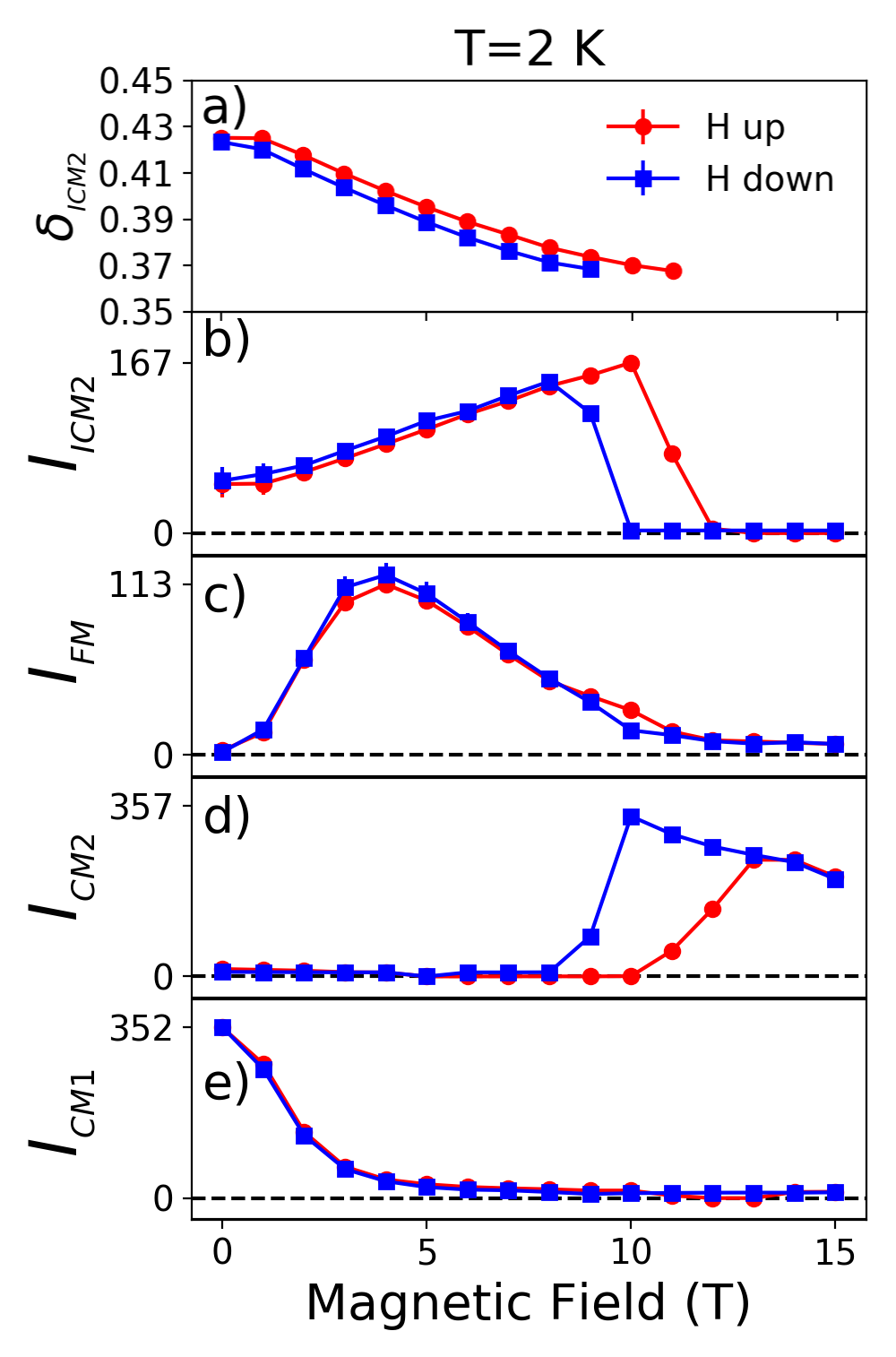}
	\caption{Evolution as a function of the magnetic field along the $b$ axis and T=2 K of the incommensurate position $\delta$ of the $q_{ICM2}$=$(\frac{3}{2},0,\delta)$ phase (a), its associated intensity $I_{ICM2}$ (b), the intensity of the ferromagnetic contribution $I_{FM}$ (c), the intensity $I_{CM2}$ at $q_{CM2}$=$(\frac{3}{2},0,\frac{1}{4})$ (d) and the $q_{CM1}$=$(\frac{3}{2},0,0)$ intensity $I_{CM1}$.}
	\label{Hdep_IP}
\end{figure}

Overall, and apart from the appearance of the ferromagnetic component, the effect of the field is to restore the $q_{CM2}=(3/2,0,1/4)$ magnetic phase, common to the other members of the RMn$_2$O$_5$ family. Here, such cycloidal structure wraps around the $c$ axis, allowing to align 2 spins out of 4 in a direction perpendicular to the field, which limits the Zeeman energy and satisfies the antiferromagnetic exchanges. The obtained phase diagram is also similar to the one derived from magnetization and dielectric constant measurements on this compound \cite{Bukhari2016}, with comparable number of phases and consistent phase boundaries in the field-temperature diagram. Details of the measured temperature and field evolution of those different Bragg reflections are gathered in Fig. \ref{Maps} and \ref{Hdep_IP}. Fig. \ref{Maps} especially shows Q-scans concatenated to produce maps as a function of wavevector and either temperature or field. Fig \ref{Hdep_IP} shows the magnetic field dependence of the intensity and/or q-position determined from a fit of those scans. \\

%===========================================================================================================================
%\begin{figure}[h]
%	\includegraphics [width=0.8\linewidth, angle=0]{Gd_0T}
%	\caption{Magnetic structure of the commensurate magnetic order in the ferroelectric phase of \gd at zero magnetic field. Symmetry elements are represented, together with relevant exchange interactions J$_3$, J$_4$ and J$_6$}
%	\label{Gd_0T}
%\end{figure}  
%===========================================================================================================================

% Magnetic-temperature phase diagram from neutron
%===========================================================================================================================
%===========================================================================================================================

\begin{figure}[ht]
	\includegraphics [width=0.99\linewidth, angle=0]{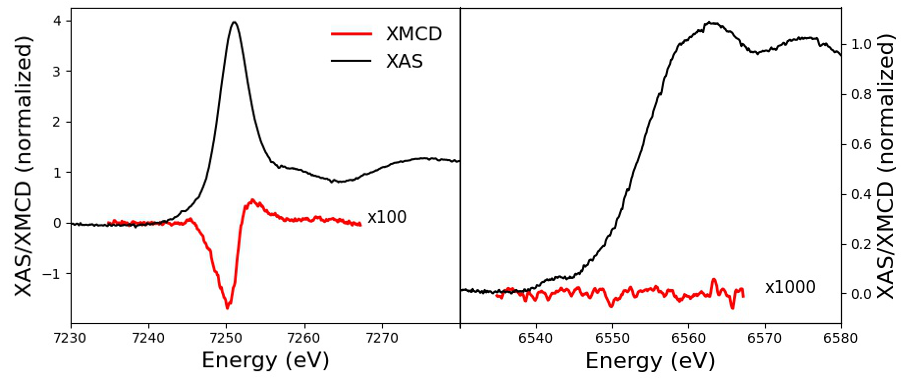}
	\caption{X-ray Absorption Spectra and associated dichroic signal at Gd (left) and Mn (right) edge, performed at 1.3 T and 6 K. Dichroic signals are multiplied by a factor 100 and 1000 respectively for a sake of visibility.}
	\label{XMCD}
\end{figure}

{\bf Evolution at 2K under magnetic field. } At 2 K, a contribution appears at $q_{FM}$=(1,0,0), where nuclear contribution is forbidden by the average structure $Pbam$ (see Fig \ref{Maps}a). As represented in Fig. \ref{Hdep_IP}c, the intensity of this peak increases and reaches a maximum for H=4 T and decreases progressively until 12 T, where it stabilizes at a minimum value up to 15 T. Fig. \ref{Maps}e-h shows that the transition temperature increases from 15 K at 1 T to more than 45 K at 15 T. As it preserves the lattice translation symmetry, it can be interpreted as a ferromagnetic contribution emerging progressively under magnetic field. Owing to the absence of magnetic anisotropy at the Gd site, contrary to the Mn site, and since the coupling between the Gd and Mn moments is the weakest in this compound \cite{Vaunat2021}, it is very likely that Gd is the main contributor to this signal. To confirm this interpretation, we performed X-ray Magnetic Circular Dichroism on a \gd\ powder sample at the ODE beamline, synchrotron SOLEIL, at both Mn K edge and Gd L$_3$ edge \cite{Baudelet2011,Baudelet2016}. As reported in Fig. \ref{XMCD}, the measurement at 1.3 T and 6 K exhibits a clear ferromagnetic contribution from the Gd but no sizeable contribution from the Mn moments with experimental sensitivity. By comparing with the literature \cite{Mito2009, Sorg2007, Ramos2013}, we were able to estimate the magnitude of the moment to be about 2.5 $\mu_B$ for Gd and below 0.05 $\mu_B$ for Mn. These results confirm the rise of a ferromagnetic component under magnetic field and further attest that this contribution is due to the Gd moments. This can easily be explained with the same mechanism reported for PrMn$_2$O$_5$ \cite{Chattopadhyay2020} replacing the exchange interaction values by the one of \gd\ \cite{Vaunat2021}.

%XMCD

Coming back to neutron scattering results at 2 K, Fig. \ref{Maps}b and Fig. \ref{Hdep_IP}e show further that the (commensurate) magnetic signal at $q_{CM1}$=(3/2,0,0) decreases rapidly up to 5 T while a very small intensity remains up to H$_C$=11 T. This $q_{CM1}$ intensity coexists at low field with the magnetic phase $q_{ICM2}$=($\frac{3}{2}$,0,$\delta$=0.425) below 4 K, as can be seen on the maps Fig. \ref{Maps}b and \ref{Maps}i. The evolution of the intensity and position of $q_{ICM2}$ is reported in Fig. \ref{Hdep_IP}b and Fig. \ref{Hdep_IP}a respectively. From 0 to 10 T, the intensity grows and the position shifts from $\delta$=0.425 to $\delta$=0.37. Between 0 and 1 T the position remains the same, suggesting a possibility that the 0.425 position may correspond to a commensurate order with q=3/7$\approx0.428$. At H$_C$=11 T a magnetic transition occurs, the $q_{ICM2}$ signal disappears, and another magnetic signal appears at $q_{CM2}$=($\frac{3}{4}$,0,$\frac{1}{4}$). This critical value of magnetic field is similar to the observed transition in both magnetization measurements and electric polarization for field along $b$ direction. \\

%high temperature
{\bf Evolution above 33 K under magnetic field. } The evolution of the high temperature magnetic order as function of the field was also investigated. At 35 K, Fig. \ref{Maps}c shows that the magnetic peak $q_{ICM1}$=($\frac{3}{2}$,0, $\epsilon$) shifts from $\epsilon$=0.17 to $\epsilon$=0.22 with a decreasing intensity from 0 to 12 T. Interestingly, a commensurate peak appears above H=5 T at $q_{CM2}$=($\frac{3}{2}$,0,$\frac{1}{4}$), the very same position as the high field low temperature one, showing that this new magnetic order stabilizes at lower field at high temperature. Performing the same study at 40 K shows no sign of the commensurate phase at $q_{CM2}$, while $\epsilon$ moves from 0.19 to 0.22 between 0 and 15 T (see Fig. \ref{Maps}d).\\

{\bf Temperature evolution of field induced magnetic phases.} Despite a strong reduction of the intensity of the ferromagnetic contribution at $q_{FM}$ above 4 T, the temperature range of existence of such contribution increases from below 10 K at 1 T up to 40 K at 15 T (Fig. \ref{Maps}e-h). However, the transition temperature of $q_{CM1}$ slowly decreases from 33 K at zero field down to 25 K at 10 T (Fig. \ref{Maps}i-j). On the contrary, $q_{ICM2}$ shows an increase of its transition temperature from 3.75 to 40 K between 1 and 15 T (Fig. \ref{Maps}j-m). Both disappear above the transition field H$_C$=11 T. At 10 T however, $q_{CM2}$ appears in the narrow temperature range between 25 and 40 K. Above 12 T, this phase is present below 35 K down to 2 K (Fig. \ref{Maps}k). An additional subtlety appears at 12 T around this $q_{CM2}$ magnetic peak. Indeed, two satellite peaks appear at $q_{ICM3}$=(1.5,0,0.2) and (1.5,0,0.3) from 2 to 15 K with a maximum at 10 K (Fig. \ref{Maps}l). These satellites are still present at 15 T but disappear above 10 K. Further investigations are necessary to better identify the corresponding magnetic structures but we anticipate that these satellites may be the signatures of some kind of roughening of the cycloid around the $c$-axis or discommensurations.\\

{\bf Hysteretic behaviour.} In order to investigate the nature of the transition at H$_C$=11 T, we performed the same measurement while decreasing the field. Fig. \ref{Hdep_IP} shows the field evolution of these peaks. A clear hysteresis is visible on the intensity of $q_{ICM2}$ (\ref{Hdep_IP}b) and $q_{CM2}$ (\ref{Hdep_IP}d) with a width around 2 T. This behavior, together with the coexistence of both orders between 10 and 12 T, strongly suggests a first order transition.

\section{Discussion}
A similar hysteresis in electric polarization as a function of magnetic field was reported on the same \gd\ compounds \cite{Ponet2022}, with the difference that the magnetic field was oriented 10 degrees off the $a$ direction in the $(a,b)$ plane, with a hysteresis effect on the polarization along $b$. The authors interpreted this result arguing the presence of 4 topological states, that can be manipulated by the orientation of the applied magnetic field. The authors further relate these 4 states to 4 different spin configurations within the same zero-field unit cell, doubled along the $a$ direction, corresponding to the $q_{CM1}$ propagation wave vector. The orientation of the magnetic field in the $(a,b)$ plane seems crucial and the angle between the field and $a$-axis very specific to obtain such topological transition. The proposed analysis is based on a simple model consisting in two spin chains per unit cell, taking into account the anisotropy and only two exchange interactions (intra-chain along $a$ and inter-chains along $b$). This simple model is also based on the assumption that the intra-chain exchange interaction is dominant in as observed in most RMn$_2$O$_5$.\\

Although this simplified model allows to introduce the interesting topological aspects of these different ground states, the present results show that the model lacks some ingredients to capture the physics of the high magnetic field phase. First we show here that the topological transition can also be observed for a totally different field direction (here along $b$), with an effect on the polarization along $a$. Secondly, inelastic neutron scattering experiments single out \gd\ as an exception among the RMn$_2$O$_5$ family, with a weak intra-chain coupling \cite{Vaunat2021} thus not dominant. Finally, the present results show the importance of the coupling along the $c$ axis and call for a 3D model. Indeed, i) the magnetic transition is accompanied by a quadrupling of the unit cell in the $c$ direction (from $q_{CM1}$ to $q_{CM2}$) and ii) the anomaly at H$_c$ is systematically observed in susceptibility and polarization measurements,  whatever the direction of the magnetic field, which indicates a change of magnetic order whatever the direction of the magnetic field. \\
%interpretation
%===========================================================================================================================
\begin{figure}[ht]
	\includegraphics [width=0.9\linewidth, angle=0]{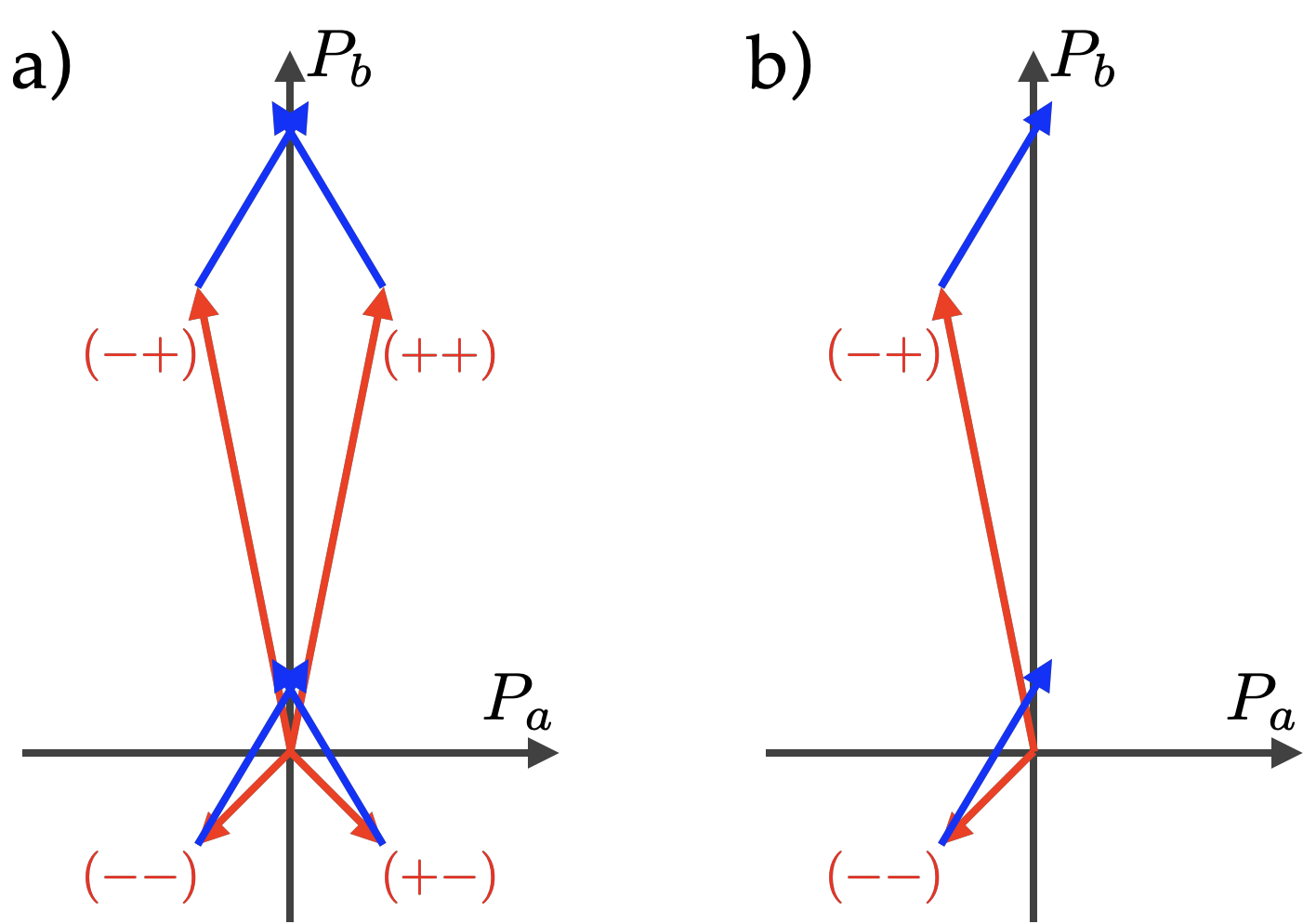}
	\caption{a) Schematic representation of the electric polarization contribution from all degenerate states at room temperature (red arrows) taken from ref \cite{Dai2020}. Blue arrow correspond to the additional spin induced contribution at low temperature. Summing all contribution lead to a b-component only polarization. b) Proposed remaining states after an hysteretic cycle with magnetic field along b above H$_c$=11 T. While the $b$ component along $b$ remain unchanged, a finite component appears along $a$.}
	\label{Interpretation}
\end{figure}  
%===========================================================================================================================
We propose a simpler reasoning based on DFT calculations of the electronic ground state \cite{Dai2020} at zero field. According to the symmetry of the room temperature $Pm$ space group, the two components of polarization along $a$ ($P_a$) and $b$ ($P_b$) are allowed. This results in 4 degenerate ``high temperature'' configurations depending on the relative sign of these components : $(++), (+-), (-+)$ and $(- -)$ as represented by the red arrows in Fig. \ref{Interpretation}a. These four configurations are associated with two different spin structures that are equivalent from a symmetry point of view and cannot be discriminated. Interestingly, these two spin configurations correspond to the two topological phases proposed at low field in Ref  \cite{Ponet2022}. According to DFT calculations \cite{Dai2020}, a spin-induced contribution adds up below the magnetic transition temperature (T$_N$=33 K) : $-\Delta P_a$ and $+\Delta P_b$ for the $(++)$ and $(+-)$ states, and $+\Delta P_a$ and $+\Delta P_b$ for the $(-+)$ and $(- -)$ states (see blue arrows in Fig. \ref{Interpretation}a). The net resulting polarization below T$_N$, summing over the 4 states, is thus along $b$ only, as measured experimentally. Above 12 T, the $q_{CM2}$ magnetic state induces a new electronic ground state as revealed experimentally by the changes in the $P_a$ and $P_b$ component of the polarization. 
In an analogous way to the proposed switching model between topological states\cite{Ponet2022}, one can expect the magnetic field to select only certain states once it has returned to zero. In order to reproduce the experimental observations, only two possible choices for the final states seem possible : $(++) + (+-)$ or $(- -) + (-+)$ (see Fig. \ref{Interpretation}b).
From a magnetic point of view, the initial state and the final state after field-sweep, are equivalent. This is consistent with the present magnetization and neutron scattering results. From an electronic point of view, however, the final state can be discriminated from the initial state. Indeed, while $P_b$ remains unchanged (see the $b$ component in Fig. \ref{Interpretation}.b), $P_a$ is no longer compensated by the two other contributions (coming from $(++)$ and $(+-)$). DFT calculations predicted a contribution of $\Delta P_a \approx$ 10 nC.cm$^{-1}$  \cite{Dai2020} which is in perfect agreement with our results in Fig. \ref{P_H}. The detailed mechanism to describe how the field along $b$ selects only certain electronic states remains to be unveiled.
%%%%%%%%%%%%%%%%%%%%%%%%%%%%%%%%%%%%%%%%

\section{Conclusion}
In conclusion, the present study reveals a first order magnetic transition under magnetic field in \gd. This transition recalls metamagnetic transitions characteristic of heavy fermions \cite{spielberg, aoki2013}. For H along $b$, the transition is characterized by a new propagation vector $q_{CM2}$=$(\frac{1}{2},0,\frac{1}{4})$, which is common among the RMn$_2$O$_5$ family. The new high field phase presents an additional contribution to the electric polarization along both $a$ and $b$ directions. Surprisingly, the polarization along $a$ does not go back to zero while decreasing the magnetic field, but remains finite. This indicates a change of the electronic ground state after magnetic field cycling, while no indication for such a change of magnetic ground state could be evidenced neither by magnetic susceptibility nor neutron diffraction. The establishment of a new electronic ground state after H cycling is further observed for all directions of the magnetic field. These results challenge previous model of magneto-electric switching between different topologically protected states. They should motivate further experimental and theoretical work to investigate the stability of these new states for other directions of the magnetic field, and also study the consequence on the dynamical properties such as the electromagnon. Our work confirms the importance of the path followed in the three dimensions (T,E,H) phase diagram in the establishment of the ground state in multiferroic materials as suggested in recent work\cite{Dai2020}, and should be extended to other external parameters such as pressure.

\section{Acknowledgments}

We thank W. Knafo for fruitfull discussions. This study was supported by grants from LLB and SOLEIL synchrotron, and LabEx PALM through Contract No. ANR-10-LABX-0039-PALM. Experiments at ILL were sponsored by the French Neutron Federation (2FDN), with Data References 10.5291/ILL-DATA.CRG-2839 and 10.5291/ILL-DATA.4-01-1700. We acknowledge SOLEIL for provision of synchrotron beamtime (proposal number 20220714) on ODE beamlines and the support of the HLD at HZDR, member of the European Magnetic Field Laboratory (EMFL proposals DMA14-219 and DMA15-219) for magnetization and polarization measurements, and the MORPHEUS platform at the Laboratoire de Physique des Solides for sample alignment.

\bibliography{biblio}
\end{document}